\documentclass[11pt]{article}
\usepackage{moriond,epsfig}

\def\be{\begin{equation}}
\def\ee{\end{equation}}
\def\bea{\begin{eqnarray}}
\def\eea{\end{eqnarray}}

\def\gsim{\mathrel{\rlap{\lower4pt\hbox{\hskip1pt$\sim$}}
     \raise1pt\hbox{$>$}}}         

\begin{document}
\vspace*{4cm}
\title{  $B \to K^{(*)} \ell^+ \ell^-$ @ Low Recoil and Physics Implications  }
DO-TH 13/09,
QFET-2013-02

\author{ G. HILLER }

\address{Institut f\"ur Physik, Technische Universit\"at Dortmund, D-44221 
Dortmund, Germany}

\maketitle\abstracts{ 
This talk covers recent theoretical progress in exclusive semileptonic rare  $B$-decays at  low hadronic recoil. The efficient
parametric suppression of the $1/m_b$ corrections in this region provides opportunities to probe the Standard Model and beyond at precision level. Notably, angular analysis allows to simultaneously access
electroweak flavor physics {\it and} hadronic matrix elements, the latter of which constitute the 
leading source of theoretical uncertainty.
Ratios of $B \to K^*$ form factors can already be extracted from present data.
A comparison  with existing theoretical determinations by lattice QCD and light cone sum rules gives a consistent picture over the whole kinematic range. In the future improved analyses will advance  our understanding of  non-perturbative methods for QCD {\it and} of $|\Delta B|=1$ transitions.}

\section{Introduction}

At the time of the writing of this article,  several flavor changing neutral current (FCNC)
exclusive semileptonic   $|\Delta B|=|\Delta S|=1$ transitions have been measured:
\begin{itemize}
\item $B \to K^{(*)} \mu^+ \mu^-$  by    BaBar \cite{BabarLakeLouise}, Belle \cite{:2009zv}, CDF \cite{HidekiICHEP2012} (with the whole CDF data sample of 9.6 fb$^{-1}$), LHCb \cite{Aaij:2013iag} (with 1 fb$^{-1}$),
Atlas \cite{ATLAS:2013ola} and  CMS \cite{CMS:cwa}.

\item $B_s \to \Phi \mu^+ \mu^-$    by  CDF  \cite{HidekiICHEP2012}, LHCb \cite{Aaij:2013aln}.

\item $\Lambda_b \to \Lambda \mu^+ \mu^-$   by   CDF \cite{HidekiICHEP2012}, LHCb \cite{LHCb-PAPER-2013-025}.
\end{itemize}
The enormous increase in event rates has made branching ratios 
and many beyond-basic observables, such as CP and forward-backward asymmetries and
other angular distributions in several dilepton mass bins  available.
The dominant fraction of  the data  originates from hadron collider experiments, 
and consists of $\mu^+ \mu^-$ final states. Note, however LHCb's  study   at low dielectron mass \cite{Aaij:2013hha}.

Different theory descriptions apply to the kinematic regions of large and low hadronic recoil.
The former corresponds to small dilepton masses $\sqrt{q^2}$ below the mass of the $J/\Psi$ 
and is accessible to QCD factorization \cite{Beneke:2001at}.
The low recoil region corresponds to large $\sqrt{q^2} \sim {\cal{O}}(m_b)$, above the $\Psi^\prime$ mass,  and is treated with an operator product expansion (OPE) in $1/Q$, $Q=\{m_b, \sqrt{q^2}\}$   put forward by
 \cite{Grinstein:2004vb}, for recent works, see \cite{Beylich:2011aq}; for earlier studies on inclusive decays, see \cite{Buchalla:1998mt}.
 Recent research directions include optimizing the physics reach of the 
 $B\to K  \ell^+ \ell^-$ \cite{Bobeth:2007dw} and $B\to K^*( \to K \pi) \ell^+ \ell^-$  \cite{Kruger:1999xa} angular distributions  and working out the physics reach at low recoil  \cite{Bobeth:2010wg} \cite{Bobeth:2011nj} \cite{Bobeth:2012vn}. 
While the phenomenology of the low $q^2$ region has received lots of attention in the past see 
{\it e.g.}   \cite{Bobeth:2007dw} \cite{Kruger:2005ep}  \cite{Bobeth:2008ij} \cite{Egede:2008uy} \cite{Altmannshofer:2008dz} \cite{Lunghi:2010tr} \cite{Becirevic:2011bp} \cite{Das:2012kz} \cite{Descotes-Genon:2013vna} \cite{JavierMoriond},
 the theoretical activities at low recoil are rather recent.
 
 The angular distribution and its theory  in $B_s \to \Phi (\to K K) \ell^+ \ell^-$ is analogous to $B\to K^*( \to K \pi) \ell^+ \ell^-$ decays. For the former decay, however, effects from meson mixing  are larger and could be of relevance  \cite{Bobeth:2008ij}. $b$-baryon decays are currently not as developed theoretically as the meson ones. We note the unique properties regarding $\Lambda_b$ and $\Lambda$ polarization studies \cite{Hiller:2007ur}.
 We stress that the theoretical low recoil features are shared by all lepton flavors, and kinematic lepton mass effects in this region are of order
  $m_\ell^2/m_b^2$. Additional flavor splittings can be due to  lepton flavor violating contributions to $b \to s \ell^+ \ell^-$ transitions  \cite{Bobeth:2007dw} \cite{Bobeth:2012vn}.

We briefly review  the beneficial properties of the low recoil region in Section \ref{sec:loreco}. 
Extractions of  hadronic form factor ratios from low recoil data \cite{Hambrock:2012dg} \cite{Hambrock:2013} are presented in Section \ref{sec:FF}.
Note the recent developments in lattice estimations of  $B \to K^{(*)}$ form factors  
\cite{whitepaper} \cite{Liu:2011raa}, which directly apply in this regime.
 The improved New Physics reach in selected low recoil $B\to K^{*} \ell^+ \ell^-$ observables   \cite{Bobeth:2012vn} is  demonstrated  in Section \ref{sec:NP}. We conclude in Section \ref{sec:conclusions}.

\section{Benefits of low recoil \label{sec:loreco}}

When combined with  the improved Isgur-Wise heavy quark form factor relations the OPE  \cite{Grinstein:2004vb} predicts a simple structure for the $B \to K^* \ell^+ \ell^-$ transversity amplitudes 
in terms of universal short-distance coefficients $C^{L,R}$ and form factors $f_i$ as
\begin{equation} \label{eq:universal}
 A_{i}^{L,R} (q^2) \propto C^{L,R}(q^2) \ \cdot f_i (q^2) + {\cal{O}}(\alpha_s \Lambda/m_b,[ {\cal{C}}_7/{\cal{C}}_9] \Lambda/m_b) ,~~~~ i=\perp,||,0 . 
\end{equation}
As can be seen, the
power corrections receive additional parametric suppressions, and are below the few percent level. 
Eq.~(\ref{eq:universal}) has powerful phenomenological consequences \cite{Bobeth:2010wg} \cite{Bobeth:2012vn}. It 
allows to define new observables which do not
 depend on form factors \footnote{Note that this  holds in the most general
 dimension six $\bar s b \bar \ell \ell$ operator basis except for $H_T^{(2)}$ which becomes form factor dependent in the presence of scalar contributions \cite{Bobeth:2012vn}.},  as 
\begin{eqnarray}
  H_T^{(1)} (q^2) & \equiv &
  \frac{\sqrt{2} J_4}{\sqrt{- J_2^c \left(2 J_2^s - J_3\right)}} ,
  \label{eq:def:HT1}
\\
  H_T^{(2)}(q^2)  & \equiv &
   \frac{\beta_l J_5}{\sqrt{-2 J_2^c \left(2 J_2^s + J_3\right)}} ,
  \label{eq:def:HT2} \\
  H_T^{(3)} (q^2) & \equiv &
  \frac{\beta_l J_{6s}}{2 \sqrt{(2 J_2^s)^2 - J_3^2}} ,
  \label{eq:def:HT3} \\
   \label{eq:HT45}  
  H_T^{(4)}(q^2) &  \equiv &
  \frac{2  J_8}{\sqrt{-2 J_{2c} \left(2 J_{2s} + J_3 \right)}} ,
\\
  H_T^{(5)} (q^2) &  \equiv &
  \frac{- J_9}{\sqrt{(2 J_{2s})^2 - J_3^2}}   ,
\end{eqnarray} 
and corresponding CP asymmetries. Here $\beta_l=\sqrt{1-4 m_\ell^2/q^2}$.
The angular coefficients $J_k =J_k(q^2)$ can be
extracted from the $B \to K^*( \to K \pi) \ell^+ \ell^-$ angular distribution. 
Briefly, $H_T^{(4,5)}$ are SM null tests and sensitive to right-handed FCNCs, whereas $H_T^{(2,3)}$ probe similar  flavor couplings as
the forward-backward asymmetry $A_{\rm FB} \propto  [J_{6s}+J_{6c}/2]/[d \Gamma/dq^2]$, however, with much reduced hadronic uncertainties, see Section \ref{sec:NP}. 

While present data are consistent with the SM and the low recoil framework, the machinery behind Eq.~(\ref{eq:universal})  can be cross-checked model-independently. To do so we introduce the (complex-valued) parameters $\epsilon_i$, $i=\perp,||,0$, and write  $A^{L,R}_i  \propto C^{L,R}\, f_{i} (1+ \epsilon_i)$. As indicated in Eq.~(\ref{eq:universal}), the generic size of the $\epsilon_i$ from the in principle known power corrections is  a few percent. While these next-to-leading order 
$1/m_b$ corrections have been worked out \cite{Grinstein:2004vb}, only little is known presently on the additional heavy quark form factors they depend on. 
The $\epsilon_i$-ansatz allows further to quantify corrections from beyond the OPE, such as duality violating effects. Specifically, 
the size of the $\epsilon_i$  can be probed with the null tests
\begin{eqnarray} \label{eq:J7}
J_{7} &=& {\cal{O}}({\mbox{Im}}[\epsilon])+{\cal{O}}({\mbox{Im}}[C_{T(T5)} \, C_{S(P)}^*])+{\cal{O}}(m_\ell/Q \, C_{S,T5}) ,\\  \label{eq:HT1}
|H_T^{(1)}| -1&  = & {\cal{O}}({\epsilon^2}) + {\cal{O}}(\Lambda^2/Q^2 \, |C_{T(5)}|^2). 
\end{eqnarray}
As indicated,
a potential New Physics background arises in the presence of  tensor $C_{T(5)}$ and/or scalar contributions $C_{S,P}$. The best reach is found to be in $J_7$, which  can probe $\epsilon_i$ at the few percent 
level. If tensors are neglected $|H_T^{(1)}| -1$ becomes comparable or better
\cite{Bobeth:2012vn}.

Presently, none of the $H_T^{(i)}$ has been measured yet,  however, once more information on the angular distribution becomes available they are very useful observables
due to their high sensitivity to the $|\Delta B|=1$ transitions. All $H_T^{(i)}$ maintain good sensitivity at large recoil \cite{Descotes-Genon:2013vna}.

Eq.~(\ref{eq:universal})  allows further to identify observables which are
  independent of short-distance coefficients. 
  These include the  already measured 
 fraction of longitudinally polarized $K^*$, $F_L$, and the 
transverse asymmetry $A_T^{(2)}$ \cite{Kruger:2005ep}.  
Both can be used  to extract hadronic matrix elements from 
$B \to K^*\ell^+ \ell^-$ data \cite{Hambrock:2012dg}, see  Section \ref{sec:FF}.
Note that the short-distance independence of $F_L$ and $A_T^{(2)}$ at low recoil would be spoiled by beyond the SM 
right-handed FCNCs. However, at  low $q^2$ $A_T^{(2)}$ is 
 highly receptive to the latter, while being a SM null test.
Since data on $A_T^{(2)}$  are consistent
with the SM and there is no other observation of such effects we neglect them here.
Improved data and global analyses will reveal whether such effects are contributing to $b \to s$ transitions.
 
\section{$B \to K^*$ form factors data versus theory \label{sec:FF}}

\begin{figure}
\begin{center}
\psfig{figure=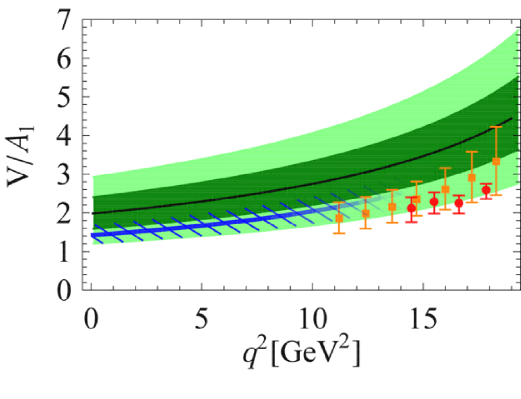,height=1.7in}
\end{center}
\caption{The data-extracted form factor ratio $V/A_1$ at $1 \sigma$ (dark) and $2 \sigma$
(lighter colored band). The data points and the hatched region denote lattice  $^{32 \, 33}$
and LCSR $^{34}$
estimations, respectively, see text.
Figure taken from $^{29}$.
\label{fig:VoverA1}}
\end{figure}

At low recoil the transversity amplitudes $A_{i}^{L,R} $, $i=\perp,||,0$ are related, see Eq.~(\ref{eq:universal}). It follows that the short-distance coefficients
$C^{L,R}$ drop out in specific observables:
\begin{eqnarray}
  F_{L} (q^2) = &
  \frac{|A_0^L|^2 + |A_0^R|^2}{\sum_{X=L,R} ( |A_0^X|^2+ |A_\perp^X|^2+|A_\parallel^X|^2 )}
 =   \frac{f_0^2}{f_0^2 + f_\perp^2 + f_\parallel^2} ,\\
    A_T^{(2)}(q^2)  = &
  \frac{|A_\perp^L|^2 + |A_\perp^R|^2-|A_\parallel^L|^2-|A_\parallel^R|^2}
       {|A_\perp^L|^2 + |A_\perp^R|^2+|A_\parallel^L|^2+|A_\parallel^R|^2}\, 
= \frac{f_\perp^2 - f_\parallel^2}{f_\perp^2 + f_\parallel^2} ,
\end{eqnarray}
and form factor ratios $f_i/f_j$ can be extracted from the corresponding data.
For other observables with this property see \cite{Bobeth:2012vn}. 
Last year first measurements of $F_L$ and $A_T^{(2)}$ have become available,  leading to first data-driven fits of $B \to K^*$ form factor ratios over the whole kinematic region \cite{Hambrock:2012dg}, see Fig.~\ref{fig:VoverA1}.
Here, the lattice and light cone sum rules (LCSR) results are shown for comparison only (they are not included in the fit) and stem from  \cite{Liu:2011raa}  \cite{Becirevic:2006nm} and  \cite{Ball:2004rg}, respectively.
The $q^2$-shape in the fit is based on a series expansion in $z=z(q^2,t_0)$ where $|z|\leq 1$ as
\begin{equation} \label{eq:series}
 f_i
(q^2) 
\propto  \frac{(\sqrt{-
z(q^2,0)
})^m
(\sqrt{z(q^2,t_-)})^l }{z(q^2,m_{R_i}^2)}
 \, \sum_{k} \alpha_{i,k} \, z^k,  ~~~~~z(t,t_0)
= \frac{ \sqrt{t_+-t}- \sqrt{t_+-t_0}}{\sqrt{t_+-t}+ \sqrt{t_+-t_0}}.
\end{equation}
Here  $t_\pm=(m_B \pm m_{K^*})^2$ and 
$l=1,0,0$ and $m=1,1,0$ for $i=\perp,\parallel,0$, respectively. The latter adjusts  the  kinematic behaviour,  $f_\perp \propto \sqrt{q^2 \hat{\lambda}} V$,
 $ f_\parallel \propto \sqrt{q^2} A_1$ etc.~in terms of the standard form factors $V,A_{1,2}$.
The factor $z(q^2,m_{R_i}^2)^{-1}$ corresponds to the off-shell poles from
axial $m_R=5.83$ GeV  ($i=\parallel, 0$)  and vector  $m_R=5.42$ GeV ($i=\perp$) $B_s^*$ mesons, respectively.  
The best-fit result in the lowest order series expansion (SE) is obtained as
$	\alpha_\parallel/\alpha_\perp=0.43^{+0.11}_{-0.08},
~
\alpha_0/\alpha_\perp=0.15^{+0.03}_{-0.02}
$ \cite{Hambrock:2012dg}. The $t_0$-dependence drops out at this order.
\begin{table}[t]
\centering
\caption{Form factor ratios at 
$q^2=0$ from $B \to K^* \mu^+ \mu^-$ data at low recoil versus theory estimations, see text.
The fits are to the lowest order series expansion (SE). The (preliminary) result SE2013 $^{30}$   
 includes the most recent data $^{ 3\, 4 \, 5 \, 6}$
and supersedes SE. Table adapted from $^{29}$,
see text. $^*$Errors symmetrized and added in quadrature. 
	\label{tab:ratios}}
\vspace{0.4cm}
\begin{tabular}{|c|ccccc|} 
\hline
\mbox{}
& 
SE & SE2013 & LCSR$^*$~\cite{Ball:2004rg} & LCSR$^*$~\cite{Khodjamirian:2010vf}
& LEL
\\
\hline
$V(0)/A_1(0)$ 
& $2.0 \pm 0.4$&   
$1.6 \pm 0.3$ 
& 
$1.4 \pm 0.2$ & 
$1.5 \pm 0.9$ & $1.3 + {\cal{O}}(1/m_b)$
\\
$A_1(0)/A_2(0)$  & 
$1.2 \pm 0.1$ &
$1.1 \pm 0.0 $ & 
$1.1\pm 0.2$ &
$1.0 \pm 0.7$ & -
\\
\hline
\end{tabular}
\end{table}

The fitted form factor ratios at $q^2=0$ along with LCSR predictions \cite{Ball:2004rg,Khodjamirian:2010vf} are compiled in Table \ref{tab:ratios}.  The preliminary fit SE2013    \cite{Hambrock:2013} includes the most recent data 
 \cite{HidekiICHEP2012} \cite{Aaij:2013iag} \cite{ATLAS:2013ola}   \cite{CMS:cwa} and supersedes the SE fit. One observes that the fitted value of $V/A_1$ is somewhat larger than the LCSR ones and larger than the symmetry-based prediction in the large energy limit (LEL) \cite{Charles:1998dr}, 
 \begin{equation} \label{eq:LEET}
V(q^2)/A_1(q^2) =(m_B+m_{K^*})^2/(m_B^2+m_{K^*}^2-q^2) ,
\end{equation}
which receives no $\alpha_s$-corrections at leading power~\cite{Burdman:2000ku},
{\it i.e.,} $V(0)/A_1(0) = 1.3 +{\cal{O}}(1/m_b)$. This tendency has decreased with the new 2013 data, but even before, in the SE fit,  there was agreement within $\sim 1 \sigma$.

Overall, the data-extracted form factor ratios  and their extrapolations are found to be currently
consistent with theoretical estimations by different non-perturbative means applicable to different $q^2$-domains. The fits at low recoil provide direct experimental information on form factor ratios in this region; this is useful for  lattice estimations  as a future benchmark test. Extrapolations to the low $q^2$ range are more sensitive to the parametrization of the slope. Fits to a more involved shape  than the lowest order series expansion require some input from large recoil to be predicitive
at large recoil \cite{Hambrock:2013}.
To proceed it would be useful if in addition to the form factors itself also their ratios could be provided  in theoretical estimations. Note, that the lattice and LCSR based ratios shown in Table \ref{tab:ratios} and
Fig.~\ref{fig:VoverA1} are obtained for lack of correlation information assuming gaussian error propagation.
 In addition, with finer binning the sensitivity to the $q^2$-shape would increase, however,
 this needs to be done while watching  the OPE.

\section{Probing New Physics \label{sec:NP}}

\begin{figure}
\begin{center}
\psfig{figure=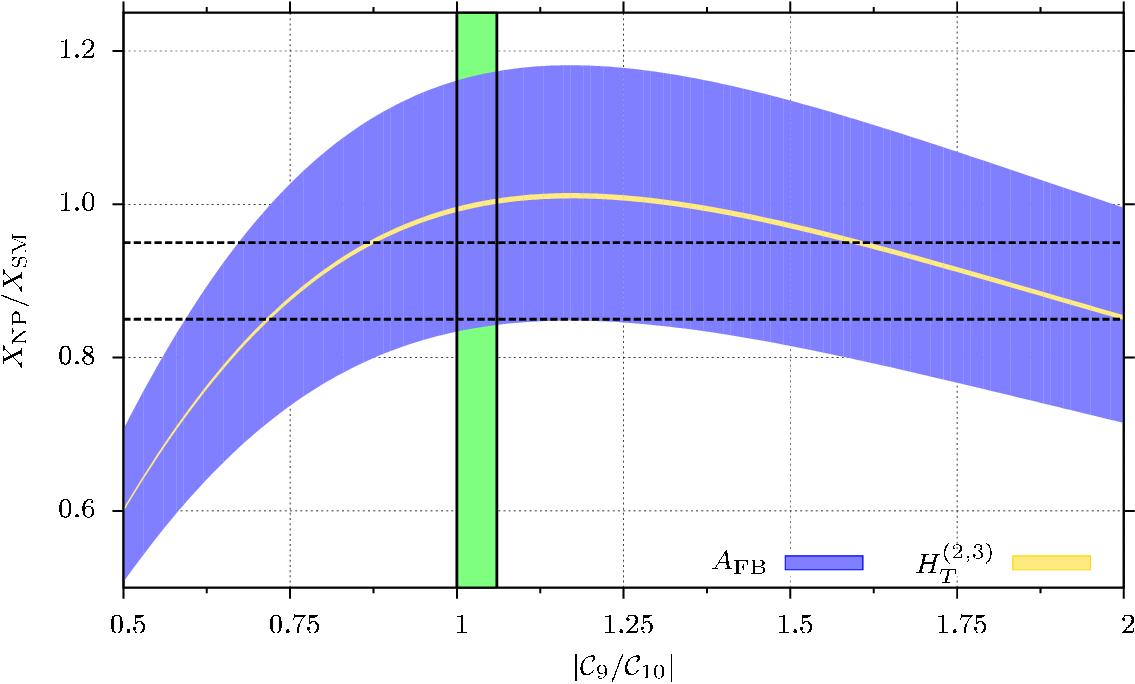,height=1.4in}
\end{center}
\caption{Sensitivity to $|{\cal{C}}_9/{\cal{C}}_{10}|$ in $A_{\rm FB}$ (blue, wide band) and $H_T^{(2,3)}$ (gold, thin band) integrated over the low recoil region $q^2 \gsim 14.2 \,{\mbox{GeV}}^2$. The vertical (green) band denotes the SM prediction. Figure taken from  $^{18}$.
\label{fig:NP-AFB}}
\end{figure}

The reach in low recoil-integrated $B \to K^* \ell^+ \ell^-$ observables versus the ratio of  Wilson coefficients $|{\cal{C}}_9/{\cal{C}}_{10}|$
is shown in Fig.~\ref{fig:NP-AFB}. The blue (broad) band corresponds to $A_{\rm FB}$, which
is form factor dependent, whereas the thin light-colored band belongs to
the form factor-free observables $H_T^{(2,3)}$. Note that $H_T^{(2)}=H_T^{(3)}$ in the SM operator basis.

\begin{figure}
\begin{center}
\psfig{figure=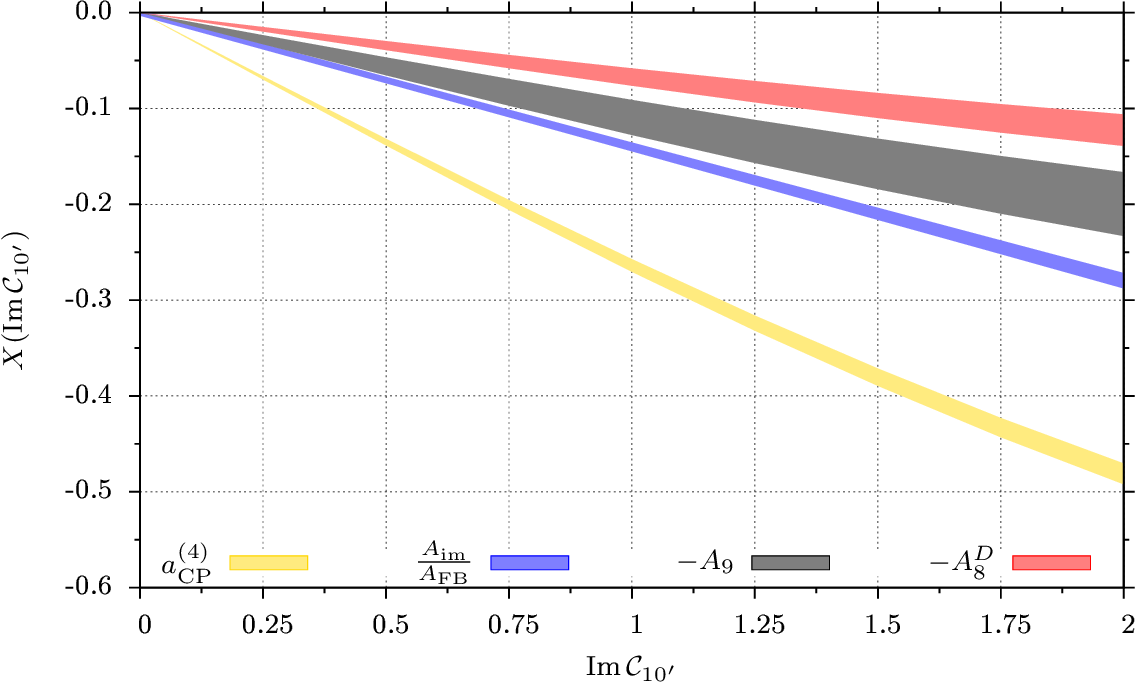,height=1.4in}
\end{center}
\caption{The sensitivity of  various $B \to K^* \ell^+ \ell^-$ CP-asymmetries to  New Physics in ${\cal{C}}_{10}^\prime$. Figure taken from  $^{18}$.
%
\label{fig:NP-ACP}}
\end{figure}

The reach in low recoil-integrated  CP-asymmetries versus the Wilson
coefficient ${\cal{C}}_{10}^\prime$ is shown in Fig.~\ref{fig:NP-ACP}. All asymmetries shown are SM null tests.
The 
gold curve with the steepest slope corresponds to the CP-asymmetry of $H^{(4)}_T$, whereas the blue curve with the next-to-steepest slope to $A_{im}/A_{\rm FB} =J_9/J_{6s} \propto H^{(5)}_T/H^{(3)}_T$, respectively. Both asymmetries are form factor-free; the other two CP asymmetries $A_{8,9}$ \cite{Bobeth:2008ij} are not, and exhibit larger uncertainties than the former.

\section{Conclusions \label{sec:conclusions}}

Smaller uncertainties in the measurements and access to further observables in exclusive $b \to s \ell \ell$ modes are expected to come from the LHC experiments in the nearer future, especially for  muons.
Belle II and a possible super flavor factory provide opportunites for other flavors in the final states as well, including inclusive decays, electrons, neutrinos and possibly taus.

For a clear interpretation of the data at precision level 
parametric (form factors, CKM, ...) and systematic inputs (framework, power corrections, ...) 
 need to be disentangled from the 
 $|\Delta B|=1$ couplings.
The important feature  in $B \to V \ell^+ \ell^-$ decays at low recoil is the 
ability to access these inputs individually.
Amended by other  information,  such as improved calculations of the hadronic matrix elements or complementary measurements this gives way to a map of the $b \to s$ flavor
structure at the weak scale; the $b \to d$ one is already at the horizon.

\vspace{-0.2cm}

\section*{Acknowledgments}
\vspace{-0.2cm}
GH is most happy to thank Christoph Bobeth and Danny van Dyk for comments on the manuscript and the organizers of the 2013 Moriond Electroweak session for the invitation to the stimulating meeting.
The works presented here are supported in part by the Bundesministerium f\"ur Bildung und
Forschung (BMBF) and by the Deutsche Forschungsgemeinschaft (DFG) within research unit
{\bf \sc FOR} 1873.

\end{document}